\documentclass[aps,pra,showpacs,twocolumn,floats]{revtex4}
\usepackage{graphicx,psfig}
\usepackage{dcolumn}
\usepackage{bm}

\input{tcilatex}

\begin{document}

\title{Nonadiabatic Transitions for a Decaying Two-Level-System: \\
Geometrical and Dynamical Contributions}
\author{R. Schilling${}^1$, Mark Vogelsberger${}^1$ and D. A. Garanin ${}^2$}
\affiliation{ ${}^1$ Institute of Physics, Johannes
Gutenberg-Universit\"at Mainz,
Staudinger Weg 7, D-55099 Mainz, Germany \\
${}^2$ \mbox{Physics Department, Lehman College, City University
of New York,} \\ \mbox{250 Bedford Park Boulevard West, Bronx, New
York 10468-1589, U.S.A.}}
\date{\today}

\begin{abstract}
We study the Landau-Zener Problem for a decaying two-level-system
described by a non-hermitean Hamiltonian, depending analytically
on time. Use of a super-adiabatic basis allows to calculate the
non-adiabatic transition probability $P$ in the slow-sweep limit,
without specifying the Hamiltonian explicitly. It is found that
$P$ consists of a ``dynamical'' and a ``geometrical'' factors. The
former is determined by the complex adiabatic eigenvalues
$E_\pm(t)$, only, whereas the latter solely requires the knowledge
of $\alpha _\pm(t)$, the ratio of the components of each of the
adiabatic eigenstates. Both factors can be split into a universal
one, depending only on the complex level crossing
points, and a nonuniversal one, involving the full time dependence of $%
E_\pm(t)$. This general result is applied to the Akulin-Schleich
model where the initial upper level is damped with damping
constant $\gamma$. For
analytic power-law sweeps we find that St\"uckelberg oscillations of $%
P$ exist for $\gamma$ smaller than a critical value $\gamma _c$
and disappear for $\gamma > \gamma _c$. A physical interpretation
of this behavior will be presented by use of a damped harmonic
oscillator.
\end{abstract}
\pacs{34.10.+x, 32.80.Bx, 03.65.Vf} \maketitle


\section{Introduction}

In many cases one can reduce the quantum behavior of a system to that of a
two-level system (TLS), which corresponds to a (pseudo-)spin one half. The
spin-down and spin-up state will be denoted by $|1 \rangle$ and $|2 \rangle$%
, respectively. If the TLS is in state $|\Psi _0 \rangle$ at time $t_0$ one
obtains $|\Psi (t) \rangle$ by solving the Schr\"odinger equation

\begin{equation}
i\hbar \frac{d}{dt}|\Psi (t)\rangle =H(t)|\Psi (t)\rangle  \label{eq1}
\end{equation}
with initial condition $|\Psi (t_{0})\rangle =|\Psi _{0}\rangle $. Note,
that we allow for an explicit time dependence of $H$. One of the quantities
of particular interest is the survival probability
\begin{equation}
P=\lim_{t\rightarrow \infty }\lim_{t_{0}\rightarrow -\infty }|\langle \Psi
\left( t_{0}\right) |\Psi (t)\rangle |^{2}  \label{eq2}
\end{equation}
that the system remains in its initial state. For a TLS with a level spacing
depending linearly on time the result for $P$ as function of the sweep rate $%
v$ has been derived approximately by Landau \cite{1} and St\"{u}ckelberg
\cite{2} and rigorously by Zener \cite{3} and Majorana \cite{4}. $P$ will
depend sensitively on the $t$-dependence of $H$ and can not be calculated
analytically, except in limiting cases, only. One of them is the adiabatic
limit. In that limit it is known that $|\Psi (t)\rangle $ converges to a
superposition of the adiabatic states $|u_{0,\pm }(t)\rangle $ which are
solutions of the eigenvalue equation:
\begin{equation}
H(t)|u_{0,\pm }(t)\rangle =E_{\pm }(t)|u_{0,\pm }(t)\rangle  \label{eq3}
\end{equation}
with $E_{\pm }(t)$ the adiabatic eigenvalues. Although $E_{+}(t)$ and $%
E_{-}(t)$ may not cross in real time (avoided level-crossing) this will
happen for complex times $t_{c}^{k},\;k=1,2,\ldots ,N$.

In case of a \textit{real-symmetric} Hamiltonian matrix $\langle \nu
|H(t)|\nu ^{\prime }\rangle ,\,$\ $\ \,\,\nu ,\nu ^{\prime }=1,2$ which is
analytic in $t$ and for a single crossing point $t_{c}$ in the upper complex
$t$-plane ($\mathrm{Im}t_{c}>0$) it was shown by Dykhne \cite{5} (see also
earlier work by Pokrovskii et al. \cite{6}) that
\begin{equation}
P\cong \exp [-2\;\mathrm{Im}\;z(t_{c})]  \label{eq4}
\end{equation}
in the adiabatic limit. The new variable $z(t)$ is given by
\begin{equation}
z(t)=\int\limits_{0}^{t}dt^{\prime }[E_{+}(t^{\prime })-E_{-}(t^{\prime })].
\label{eq5}
\end{equation}

Davis and Pechukas \cite{7} have performed an exact proof of result (\ref
{eq4}), (\ref{eq5}). Particularly, these authors have proven that the
pre-exponential factor equals one. Therefore it is sometimes called the
Dykhne-Davis-Pechukas (DDP) formula. For more than one crossing point with $%
\mathrm{Im}z_{c}^{k}=\mathrm{Im}z(t_{c}^{k})>0$ a generalization of (\ref
{eq4}) has been suggested \cite{7,8} and tested by Suominen and co-workers
(Ref. \cite{9} and references where-in). A rigorous prove of the
generalization of DDP-formula including even \textit{hermitean} Hamiltonians
has been provided by Joye et al. \cite{10}. More than one crossing point
leads to interferences which generate oscillations in $P$ as function of
control parameters, like the sweeping rate (see below).

For Hamiltonian matrices which are not real-symmetric, but hermitean, Berry
\cite{11} and Joye et al. \cite{12} made an interesting observation which is
that $P$ obtains also a ``geometrical'' factor besides the ``dynamical''
one, Eq.~(\ref{eq4}), where the former also depends on the crossing points $%
t_c^k$, only. For those who are less familiar with this kind of physics let
us explain the choice of this nomenclature. Below we will see that one of
the factors of $P$ is entirely determined by the adiabatic eigenvalues and
the other by the adiabatic eigenstates. Since the former is important for
the time evolution it is called ``dynamical'' whereas the latter is related
to the geometry in the Hilbert space, particularly through a condition for
parallel transport (Eq.~\ref{eq21}), and accordingly it is called
``geometrical''.

TLS will be influenced by their environment, e.g. by phonons. The
spin-phonon coupling leads to dissipation of the (pseudo-)spin dynamics
which will influence the probability $P$. Although there exist microscopic
models for the spin-boson system \cite{13}, and simplified models where the
bath is described by fluctuating fields \cite{14,15,16,17,18}, we will use a
dissipative Schr\"odinger equation. This will be achieved by using a \textit{%
non-hermitean} Hamiltonian for the TLS. A particular version of such a model
has been suggested by Akulin and Schleich \cite{19}. In their model, called
AS-model in the following, the upper level (at the initial time $t_0$)
experiences a damping (see section III).

The survival and transition probability for non-hermitean TLS-Hamiltonians
has already been investigated by Moyer \cite{20}. This has been done by
mapping the original differential equation to the Weber equation, which can
be solved exactly. By use of the Weber equation as the appropriate
``comparison equation'' it was shown how the DDP-formula, Eqs. (\ref{eq4})
and (\ref{eq5}), can be extended \cite{20}. However, this extension does not
contain a ``geometric'' contribution, although one expects that it exists
similarly to what has been proven for hermitean matrices \cite{11,12}. On
the other hand Garrison and Wright \cite{21} have investigated the
geometrical phase for dissipative systems but not the non-adiabatic
transition probability.

It is one of our main goals to derive a generalized DDP-formula in the
adiabatic limit containing a ``geometrical'' and a ``dynamical''
contribution for a \textit{general} non-hermitean TLS-Hamiltonian. We will
demonstrate that both contributions consist of a universal and a
non-universal part. The former depends only on the complex crossing points
whereas the latter requires the knowledge of the complete time dependence of
$H$. Instead of using a ``comparison equation'' we apply the concept of a
\textit{superadiabatic} basis, put forward by Berry \cite{22}, to
non-hermitean TLS-Hamiltonians. As a result we will find that the
``dynamical'' contribution to the non-adiabatic transition probability
(which equals the survival probability in the adiabatic limit) is determined
by the complex, adiabatic eigenvalues $E_\pm (t)$, only. The corresponding
``geometrical'' part solely requires the knowledge of $\alpha _\pm (t)$),
the ratio of the components of each of the \textit{adiabatic} eigenstates.

A second motivation is the application of our results to the AS-model. It
has been shown that the survival probability $P$ does not depend on the
damping coefficient $\gamma$ of the upper level, provided the bias of the
TLS varies \textit{linearly} in time, and the coupling $\Delta$ between both
levels is time-independent \cite{19}. Therefore it is interesting to
investigate \textit{non-linear} time dependence and to check whether or not $%
P$ remains insensitive on $\gamma$. For nonlinear time dependence more than
one complex crossing points may occur, such that interference effects can
govern the dependence of $P$ on the sweeping rate \cite{10}. Specific
examples with $\gamma =0$ for which this happens were discussed in recent
years \cite{9,23}. There it was found that critical values for the sweeping
rate exists at which the survival probability vanishes, i.e. complete
transitions occur between both quantum levels. Consequently, one may ask:
Are these complete transitions reduced or even suppressed in the presence of
damping?

Our paper is organized as follows. The next section will contain the general
treatment of the non-hermitean Hamiltonian and the presentation of the
generalized DDP-formula. In section III we will apply the results from the
second section to the AS-model with power law time dependence. The results
for the AS-model for power law sweeps can be interpreted by the dynamics of
a damped harmonic oscillator. This will be shown in section IV. A short
summary and some conclusions are given in the final section.

\section{General Formula for Nonadiabatic Transition Probability}

In this section, we will derive a generalized DDP-formula for the
non-adiabatic transition problem of a decaying TLS. The Hamiltonian can be
represented as follows

\begin{equation}
H(\delta \,\,t/\hbar )=\frac{1}{2}\sum\limits_{j=1}^{3}B_{j}(\delta
\,\,t/\hbar )\sigma _{j}  \label{eq6}
\end{equation}
with $\sigma _{j}$, the Pauli-matrices and $B_{j}$ a time dependent field. $%
\delta >0$ is the adiabaticity parameter. Because this model should be
dissipative, at least one of the $B_{j}$ must contain a nonzero imaginary
part. Accordingly $H$ is nonhermitean. In the following we will assume that $%
B_{j}$ is analytic in $t$. Introducing a new time variable

\begin{equation}
\tau =\delta \,\,t/\hbar  \label{eq7}
\end{equation}
Eq.~(\ref{eq1}) becomes

\begin{equation}
i\delta d_{\tau }|\Psi (\tau )\rangle =H(\tau )|\Psi (\tau )\rangle ,
\label{eq8}
\end{equation}
where $d_{\tau }=\partial /\partial \tau $. $|\Psi (\tau )\rangle $ can be
expanded with respect to $|\nu \rangle $
\begin{equation}
|\Psi (\tau )\rangle =\sum\limits_{\nu =1}^{2}c_{\nu }(\tau )|\nu \rangle .
\label{eq9}
\end{equation}
With $|\Psi (\tau _{0})\rangle $, the initial state, its survival
probability is
\begin{equation}
P\equiv P(\delta )=\lim_{\tau \rightarrow \infty }\lim_{\tau _{0}\rightarrow
-\infty }|\langle \Psi (\tau _{0})|\Psi (\tau )\rangle |^{2}.  \label{eq10}
\end{equation}
Note that $P$ is the survival probability with respect to the \emph{diabatic}
basis. With respect to the \emph{adiabatic} basis $P$ is the nonadiabatic
transition probability.

To calculate $P$ for $\delta \ll 1$ we introduce the \textit{adiabatic basis}
of $H(\tau )$. This can be done as in Ref.~\cite{21} where a biorthonormal
set of right-eigenstates was used or alternatively by use of left- and
right-eigenstates. We will use the latter, as it turns out to be more
elegant. Let
\begin{equation}
|u_{0,\pm }(\tau )\rangle =\sum\limits_{\nu =1}^{2}e_{\pm }^{\nu }(\tau
)|\nu \rangle  \label{eq11}
\end{equation}
be the adiabatic \textit{right-}eigenstates. They are solutions of
\begin{equation}
H(\tau )|u_{0,\pm }(\tau )\rangle =E_{\pm }(\tau )|u_{0,\pm }(\tau )\rangle
\label{eq12}
\end{equation}
with $E_{\pm }(\tau )$, the adiabatic eigenvalues. Note that $E_{\pm }(\tau
) $ are complex in general and that the norm of $|u_{0,\pm }(\tau )\rangle $
and of $|\Psi (\tau )\rangle $ is \textit{not} conserved, since $H(\tau )$
is nonhermitean. Following Berry \cite{22}, we introduce a hierarchy of
\textit{superadiabatic} right-eigenstates $|u_{n,\pm }(\tau )\rangle $ , $%
n=0,1,2,\ldots $ and expand the solutions $|\Psi _{\pm }(\tau )\rangle $ of
Eq.~(\ref{eq8}) with respect to the superadiabatic basis:
\begin{equation}
|\Psi _{\pm }(\tau )\rangle =\exp \left[ -\frac{i}{\delta }\int\limits_{\tau
_{0}}^{\tau }d\tau ^{\prime }E_{\pm }(\tau ^{\prime })\right]
\sum\limits_{m=0}^{\infty }\delta ^{m}|u_{m,\pm }(\tau )\rangle .
\label{eq13}
\end{equation}
Substituting $|\Psi _{\pm }(\tau )\rangle $ into Eq.~(\ref{eq8}) yields the
recursion relations
\begin{equation}
\lbrack H(\tau )-E_{\sigma }(\tau )]|u_{0,\sigma }(\tau )\rangle =0\;,\quad
\sigma =\pm  \label{eq14}
\end{equation}
\begin{equation}
id_{\tau }|u_{m-1,\sigma }(\tau )\rangle =[H(\tau )-E_{\sigma }(\tau
)]|u_{m,\sigma }(\tau )\rangle \;,\;m\geq 1.  \label{eq15}
\end{equation}
Eq.~(\ref{eq14}) is already fulfilled, due to Eq.~(\ref{eq12}). To make
progress we introduce the adiabatic \textit{left}-eigenstates
\begin{equation}
\langle \tilde{u}_{0,\sigma }(\tau )|=\sum\limits_{\nu =1}^{2}\tilde{e}%
_{\sigma }^{\nu }(\tau )\langle \nu |  \label{eq16}
\end{equation}
which are solutions of
\begin{equation}
\langle \tilde{u}_{0,\sigma }(\tau )|H(\tau )=E_{\sigma }(\tau )\langle
\tilde{u}_{0,\sigma }(\tau )|  \label{eq17}
\end{equation}
and are normalized such that:
\begin{equation}
\langle \tilde{u}_{0,\sigma }(\tau )|u_{0,\sigma ^{\prime }}(\tau )\rangle
=\delta _{\sigma \sigma ^{\prime }}.  \label{eq18}
\end{equation}
Let be:
\begin{equation}
\alpha _{\sigma }(\tau )=\frac{e_{\sigma }^{2}(\tau )}{e_{\sigma }^{1}(\tau )%
},  \label{eq19}
\end{equation}
the ratio of the components of the adiabatic right-eigenstate $|u_{0,\sigma
}(\tau )\rangle $. Then it is straightforward to prove thatmm
\begin{equation}
\tilde{e}_{\sigma }^{\nu }(\tau )=\frac{\sigma }{[\alpha _{+}(\tau )-\alpha
_{-}(\tau )]e_{\sigma }^{1}(\tau )}\left\{
\begin{array}{cc}
-\alpha _{\mp }(\tau ), & \nu =1 \\
1, & \nu =2
\end{array}
\right.  \label{eq20}
\end{equation}
which defines the left-eigenstate from the right-eigenstate. Multiplication
of Eq.~(\ref{eq15}) for $m=1$ with $\langle \tilde{u}_{0,\sigma }(\tau )|$
leads to
\begin{equation}
\langle \tilde{u}_{0,\sigma }(\tau )|d_{\tau }|u_{0,\sigma }(\tau )\rangle
\equiv 0\;,\;\sigma =\pm .  \label{eq21}
\end{equation}
This is the condition for ``parallel transport'' \cite{11,24} now
generalized to nonhermitean Hamiltonians.

In order to solve recursion (\ref{eq15}) we expand $|u_{m,\sigma }(\tau
)\rangle $, $m\geq 1$ with respect to $|u_{0,\sigma }(\tau )\rangle $:
\begin{equation}
|u_{m,\sigma }(\tau )\rangle =a_{m}^{\sigma }(\tau )|u_{0,-}(\tau )\rangle
+b_{m}^{\sigma }(\tau )|u_{0,+}(\tau )\rangle .  \label{eq22}
\end{equation}
Substitution of Eq.~(\ref{eq22}) into Eq.~(\ref{eq15}) and multiplying by $%
\langle \tilde{u}_{0,\sigma }(\tau )|$ yields with Eqs.~(\ref{eq14}), (\ref
{eq18}) for $m\geq 1$:
\begin{equation}
\dot{a}_{m-1}^{-}(\tau )=-\kappa _{-}(\tau )b_{m-1}(\tau )  \label{eq23}
\end{equation}
\begin{equation}
\dot{b}_{m-1}^{-}(\tau )=-\kappa _{+}(\tau )a_{m-1}^{-}(\tau )-i[E_{+}(\tau
)-E_{-}(\tau )]b_{m}^{-}(\tau ),  \label{eq24}
\end{equation}
where $\dot{}$ denotes derivative with respect to $\tau $ and
\begin{equation}
\kappa _{\sigma }(\tau )=\langle \tilde{u}_{0,\sigma }(\tau )|d_{\tau
}|u_{0,-\sigma }(\tau )\rangle  \label{eq25}
\end{equation}
are the nonadiabatic coupling functions, responsible for the nonadiabatic
transitions. If $\kappa _{\sigma }(\tau )\equiv 0$, we get from Eqs.~(\ref
{eq23}) and (\ref{eq24})
\begin{equation}
a_{m}^{-}(\tau )\equiv a_{m}^{-}(\tau _{0})\;,\;b_{m}^{-}(\tau )=\frac{i}{%
E_{+}(\tau )-E_{-}(\tau )}\dot{b}_{m-1}^{-}(\tau ).  \label{eq26}
\end{equation}
Similar equations follow for $a_{m}^{+}(\tau ),b_{m}^{+}(\tau )$, which
however, will not be needed. Next we fix the initial condition for $|\Psi
_{\sigma }(\tau )\rangle $:
\begin{equation}
|\Psi _{\sigma }(\tau _{0})\rangle =|u_{0,\sigma }(\tau _{0})\rangle ,
\label{eq27}
\end{equation}
i.e., we start in the adiabatic right-eigenstates. From Eqs.~(\ref{eq13}), (%
\ref{eq22}) we find immediately for $\sigma =-$

\begin{eqnarray}
a_{0}^{-}(\tau ) &\equiv &1,\qquad b_{0}^{-}(\tau )\equiv 0  \nonumber
\label{eq28} \\
a_{m}^{-}(\tau _{0})=0 &,\qquad &b_{m}^{-}(\tau _{0})=0,\qquad m\geq 1.
\end{eqnarray}
such that Eq.~(\ref{eq26}) implies $a_{m}^{-}(\tau )\equiv 0$, $%
b_{m}^{-}(\tau )\equiv 0,\;m\geq 1$ provided $\kappa _{\pm }(\tau )\equiv 0$%
. This makes obvious the absence of nonadiabatic transitions.

The next step is the calculation of $\kappa _{\sigma }(\tau )$. For this we
need $e_{\sigma }^{1}(\tau )$, which can be determined from (\ref{eq21}). As
a result we find
\begin{equation}
e_{\sigma }^{1}(\tau )=e_{\sigma }^{1}(\tau _{0})\exp \left[ -\sigma
\int\limits_{\tau _{0}}^{\tau }d\tau ^{\prime }\frac{\dot{\alpha}_{\sigma
}(\tau ^{\prime })}{\alpha _{+}(\tau ^{\prime })-\alpha _{-}(\tau ^{\prime })%
}\right] ,  \label{eq29}
\end{equation}
and taking Eq.~(\ref{eq20}) into account we obtain the general result
\begin{eqnarray}
\kappa _{\sigma }(\tau ) &=&\sigma \frac{e_{-\sigma }^{1}(\tau _{0})}{%
e_{\sigma }^{1}(\tau _{0})}\exp \left[ \sigma \int\limits_{\tau
_{0}}^{0}d\tau ^{\prime }\frac{\dot{\alpha}_{+}(\tau ^{\prime })+\dot{\alpha}%
_{-}(\tau ^{\prime })}{\alpha _{+}(\tau ^{\prime })-\alpha _{-}(\tau
^{\prime })}\right]  \nonumber  \label{eq30} \\
&\times &\frac{\dot{\alpha}_{-\sigma }(\tau )}{\alpha _{+}(\tau )-\alpha
_{-}(\tau )}\exp \left[ \sigma \int\limits_{0}^{\tau }d\tau ^{\prime }\frac{%
\dot{\alpha}_{+}(\tau ^{\prime })+\dot{\alpha}_{-}(\tau ^{\prime })}{\alpha
_{+}(\tau ^{\prime })-\alpha _{-}(\tau ^{\prime })}\right] ,
\end{eqnarray}
where the expression has been split into a $\tau $-independent (first line)
and a $\tau $-dependent factor (second line). Following Berry \cite{22} we
truncate the series, Eq.~(\ref{eq13}), at the $n$-th level
\begin{equation}
|\Psi _{n,\sigma }(\tau )\rangle =\exp \left[ -\frac{i}{\delta }%
\int\limits_{\tau _{0}}^{\tau }d\tau ^{\prime }E_{\sigma }(\tau ^{\prime })%
\right] \sum\limits_{m=0}^{n}\delta ^{m}|u_{m,\sigma }(\tau )\rangle .
\label{eq31}
\end{equation}
and expand $|\Psi (\tau )\rangle $:
\begin{equation}
|\Psi (\tau )\rangle =\sum\limits_{\sigma =\pm }c_{n,\sigma }(\tau )|\Psi
_{n,\sigma }(\tau )\rangle .  \label{eq32}
\end{equation}
As initial condition we choose:
\begin{equation}
|\Psi (\tau _{0})\rangle =|\Psi _{-}(\tau _{0})\rangle ,  \label{eq33}
\end{equation}
which is equivalent to
\begin{equation}
c_{n,-}(\tau _{0})=1,\quad c_{n,+}(\tau _{0})=0,\;n\rightarrow \infty .
\label{eq34}
\end{equation}
Introducing a corresponding truncated state
\begin{equation}
\langle \tilde{\Psi}_{n,\sigma }(\tau )|=f_{n,\sigma }(\tau
)\sum\limits_{m=0}^{n}\delta ^{m}\langle \tilde{u}_{m,\sigma }(\tau )|,
\label{eq35}
\end{equation}
where the $\tau $-dependent prefactor $f_{n,\sigma }(\tau )$ has not to be
specified we obtain an equation of motion for $c_{n,\sigma }(\tau )$, after
Eq.~(\ref{eq32}) has been substituted into Eq.~(\ref{eq8}):
\begin{equation}
i\delta \dot{c}_{n,\sigma }(\tau )=\sum\limits_{\sigma ^{\prime
}}H_{n;\sigma \sigma ^{\prime }}(\tau )c_{n,\sigma ^{\prime }}(\tau )
\label{eq36}
\end{equation}
with
\begin{eqnarray}
H_{n;\sigma \sigma ^{\prime }}(\tau ) &=&\sum\limits_{\sigma ^{^{\prime
\prime }}=\pm }(\mathcal{L}_{n}^{-1}(\tau ))_{\sigma \sigma ^{\prime \prime
}}{\mathcal{H}}_{n;\sigma ^{^{\prime \prime }}\sigma ^{\prime }}(\tau )
\nonumber  \label{eq37} \\
\mathcal{L}_{n;\sigma \sigma ^{\prime }}(\tau ) &=&\langle \tilde{\Psi}%
_{n,\sigma }(\tau )|\Psi _{n,\sigma ^{\prime }}(\tau )\rangle  \nonumber \\
\mathcal{H}_{n;\sigma \sigma ^{\prime }}(\tau ) &=&\langle \tilde{\Psi}%
_{n,\sigma }(\tau )|H(\tau )-i\delta d_{\tau }|\Psi _{n,\sigma ^{\prime
}}(\tau )\rangle .
\end{eqnarray}
Eq.~(\ref{eq36}) can be rewritten as an integral equation
\begin{eqnarray}
c_{n,\sigma }(\tau ) &=&c_{n,\sigma }(\tau _{0})+\frac{i}{\delta }%
\int\limits_{\tau _{0}}^{\tau }d\tau ^{\prime }H_{n;\sigma ,-\sigma }(\tau
^{\prime })c_{n,-\sigma }(\tau ^{\prime })  \nonumber \\
&&\qquad \times \exp \left[ -\frac{i}{\delta }\int\limits_{\tau ^{\prime
}}^{\tau }d\tau ^{^{\prime \prime }}H_{n;\sigma \sigma }(\tau ^{^{\prime
\prime }})\right] .  \label{eq38}
\end{eqnarray}
Apart from the truncation, Eq.~(\ref{eq31}), the results are still exact.
Eq.~(\ref{eq38}) simplifies in the adiabatic limit $\delta \rightarrow 0$.
In leading order in $\delta $ we get from Eqs.~(\ref{eq18}), (\ref{eq31})
and (\ref{eq35})
\begin{equation}
\mathcal{L}_{n;\sigma \sigma ^{\prime }}(\tau )\cong f_{n,\sigma }(\tau
)\exp \left[ -\frac{i}{\delta }\int\limits_{\tau _{0}}^{\tau }d\tau ^{\prime
}E_{\sigma }(\tau ^{\prime })\right] \delta _{\sigma \sigma ^{\prime }}
\label{eq39}
\end{equation}
$[H(\tau )-i\delta d_{\tau }]|\Psi _{n,\sigma ^{\prime }}(\tau )\rangle $
can be found in Ref.~\cite{21}. Multiplying by $\langle \tilde{\Psi}%
_{n,\sigma }(\tau )|$ and making use of Eqs.~(\ref{eq12}), (\ref{eq18}), (%
\ref{eq22}) and (\ref{eq35}) leads to
\begin{eqnarray}
&&\mathcal{H}_{n;\sigma \sigma ^{\prime }}(\tau )=-\delta ^{n+1}f_{n,\sigma
}(\tau )\left[ E_{\sigma }(\tau )-E_{\sigma ^{\prime }}(\tau ^{\prime })%
\right]  \nonumber \\
&&\times \exp \left[ -\frac{i}{\delta }\int\limits_{\tau _{0}}^{\tau }d\tau
^{\prime }E_{\sigma ^{\prime }}(\tau ^{\prime })\right]  \nonumber \\
&&\times \left[ a_{n+1}^{\sigma ^{\prime }}(\tau )\delta _{\sigma
,-}+b_{n+1}^{\sigma ^{\prime }}(\tau )\delta _{\sigma ,+}\right] +\mathcal{O}%
(\delta ^{n+2})
\end{eqnarray}
from which follows
\begin{eqnarray}
&&H_{n;\sigma \sigma ^{\prime }}(\tau )=-\delta ^{n+1}\left[ a_{n+1}^{\sigma
^{\prime }}(\tau )\delta _{\sigma ,-}+b_{n+1}^{\sigma ^{\prime }}\delta
_{\sigma ,+}\right]  \nonumber \\
&&\times \left[ E_{\sigma }(\tau )-E_{\sigma ^{\prime }}(\tau )\right]
\nonumber \\
&&\times \exp \left\{ \frac{i}{\delta }\int\limits_{\tau _{0}}^{\tau }d\tau
^{\prime }\left[ E_{\sigma }(\tau ^{\prime })-E_{\sigma ^{\prime }}(\tau
^{\prime })\right] \right\} +\mathcal{O}(\delta ^{n+2}).
\end{eqnarray}
Note that the prefactor $f_{n,\sigma }(\tau )$ has cancelled. The diagonal
elements of $\mathbf{H}_{n}(\tau )$ are of order $\delta ^{n+2}$ and the
non-diagonal ones of order $\delta ^{n+1}$. Therefore it follows from Eqs.~(%
\ref{eq23}), (\ref{eq30}), (\ref{eq34}) and (\ref{eq38})
\begin{eqnarray}
&&c_{n,+}(\tau )\cong i\delta ^{n}\frac{e_{-}^{1}(\tau _{0})}{e_{+}^{1}(\tau
_{0})}\exp \left[ \int\limits_{\tau _{0}}^{0}d\tau ^{\prime }\frac{\dot{%
\alpha}_{+}(\tau ^{\prime })+\dot{\alpha}_{-}(\tau ^{\prime })}{\alpha
_{+}(\tau ^{\prime })-\alpha _{-}(\tau ^{\prime })}\right]  \nonumber \\
&&\times \int\limits_{\tau _{0}}^{\tau }d\tau ^{\prime }\dot{a}%
_{n+1}^{-}(\tau ^{\prime })\frac{[\alpha _{+}(\tau ^{\prime })-\alpha
_{-}(\tau ^{\prime })]}{\dot{\alpha}_{+}(\tau ^{\prime })}[E_{+}(\tau
^{\prime })-E_{-}(\tau ^{\prime })].  \nonumber \\
&&\times \exp \left[ \int\limits_{0}^{\tau ^{\prime }}d\tau ^{^{\prime
\prime }}\frac{\dot{\alpha}_{+}(\tau ^{^{\prime \prime }})+\dot{\alpha}%
_{-}(\tau ^{^{\prime \prime }})}{\alpha _{+}(\tau ^{^{\prime \prime
}})-\alpha _{-}(\tau ^{^{\prime \prime }})}\right]  \nonumber \\
&&\times \exp \left\{ \frac{i}{\delta }\int\limits_{\tau _{0}}^{\tau
^{\prime }}d\tau ^{^{\prime \prime }}[E_{+}(\tau ^{^{\prime \prime
}})-E_{-}(\tau ^{^{\prime \prime }})]\right\} .  \label{eq42}
\end{eqnarray}
The time dependence of $H(\tau )$ is chosen such that
\begin{eqnarray}
\lim_{\tau _{0}\rightarrow -\infty }|u_{0,-}(\tau _{0})\rangle =|1\rangle \;
&,&\;\lim_{\tau _{0}\rightarrow -\infty }u_{0,+}(\tau _{0})\rangle =|2\rangle
\nonumber \\
\lim_{\tau \rightarrow \infty }|u_{0,-}(\tau )\rangle \sim |2\rangle \;
&,&\;\lim_{\tau \rightarrow \infty }|u_{0,+}(\tau )\rangle \sim |1\rangle .
\label{eq43}
\end{eqnarray}
Note that the adiabatic states at initial time $\tau _{0}$ are normalized.
Since Eqs.~(\ref{eq27}), (\ref{eq33}) and (\ref{eq43}) imply
\begin{equation}
\lim_{\tau _{0}\rightarrow -\infty }|\Psi (\tau _{0})\rangle =|1\rangle
\label{eq44}
\end{equation}
we obtain from Eq.~(\ref{eq10}) for the nonadiabatic transition probability
in leading order in $\delta $
\begin{equation}
P(\delta )\cong \left| c_{n,+}(\infty )\langle 1|u_{0,+}(\infty )\rangle
\exp \left[ -\frac{i}{\delta }\int\limits_{-\infty }^{\infty }d\tau ^{\prime
}E_{+}(\tau ^{\prime })\right] \right| ^{2},  \label{eq45}
\end{equation}
where we used $\langle 1|u_{0,-}(\infty )\rangle =0$, due to Eq.~(\ref{eq43}%
). Substituting $c_{n,+}(\infty )$ from Eq.~(\ref{eq42}) with $\tau
_{0}=-\infty $ into Eq.~(\ref{eq45}) we get with Eqs.~(\ref{eq11}), (\ref
{eq29}) and $\lim_{\tau _{0}\rightarrow -\infty }e_{-}^{1}(\tau _{0})=1$
(due to Eq.~(\ref{eq43}))
\begin{eqnarray}
&&P(\delta )\cong \exp \left[ -(F_{g}^{ns}+\frac{1}{\delta }F_{d}^{ns})%
\right]  \nonumber \\
&&\times \left| \delta ^{n}\int\limits_{-\infty }^{\infty }d\tau \dot{a}%
_{n+1}^{-}(\tau )\frac{\alpha _{+}(\tau )-\alpha _{-}(\tau )}{\dot{\alpha}%
_{+}(\tau )}[E_{+}(\tau )-E_{-}(\tau )]\right.  \nonumber \\
&&\times \exp \left[ \int\limits_{0}^{\tau }d\tau ^{\prime }\frac{\dot{\alpha%
}_{+}(\tau ^{\prime })+\dot{\alpha}_{-}(\tau ^{\prime })}{\alpha _{+}(\tau
^{\prime })-\alpha _{-}(\tau ^{\prime })}\right]  \nonumber \\
&&\times \left. \exp \left\{ \frac{i}{\delta }\int\limits_{0}^{\tau }d\tau
^{\prime }[E_{+}(\tau ^{\prime })-E_{-}(\tau ^{\prime })]\right\} \right|
^{2}  \label{eq46}
\end{eqnarray}
with the \textit{nonsingular} ``geometrical'' and ``dynamical'' contribution
\begin{eqnarray}
F_{g}^{ns} &=&2\mathrm{Re}\left[ \int\limits_{0}^{\infty }d\tau \frac{\dot{%
\alpha}_{+}(\tau )}{\alpha _{+}(\tau )-\alpha _{-}(\tau )}\right.  \nonumber
\\
&&\qquad -\left. \int\limits_{-\infty }^{0}d\tau \frac{\dot{\alpha}_{-}(\tau
)}{\alpha _{+}(\tau )-\alpha _{-}(\tau )}\right]  \label{eq47}
\end{eqnarray}
and
\begin{equation}
F_{d}^{ns}=-2\mathrm{Im}\left[ \int\limits_{0}^{\infty }d\tau E_{+}(\tau
)+\int\limits_{-\infty }^{0}d\tau E_{-}(\tau )\right] ,  \label{eq48}
\end{equation}
respectively. Note that $F_{d}^{ns}=0$, for a hermitean Hamiltonian, since $%
E_{\sigma }(\tau )$ are real. The expressions for $F_{g}^{ns}$ and $%
F_{d}^{ns}$ put some constraints on $H(\tau )$, because both quantities
should be larger or equal to a constant $c>-\infty $, which requires that $%
\mathrm{Im}E_{\pm }(\tau )$ decays fast enough for $\tau \rightarrow \pm
\infty $.

The $\tau $-integral in Eq.~(\ref{eq46}) is dominated by the singularities
of $E_{+}(\tau )-E_{-}(\tau )$, for $\delta \rightarrow 0$. The adiabatic
eigenvalues and $\alpha _{\pm }(\tau )$ have the form
\begin{equation}
E_{\pm }(\tau )=\frac{1}{2}\left[ T(\tau )\pm \sqrt{T^{2}(\tau )-4D(\tau )}%
\right]  \label{eq49}
\end{equation}
\begin{equation}
\alpha _{\pm }(\tau )=\frac{-H_{11}(\tau )+H_{22}(\tau )\pm \sqrt{T^{2}(\tau
)-4D(\tau )}}{2H_{12}(\tau )},  \label{eq50}
\end{equation}
where $T$ and $D$, is respectively, the trace and the determinant of the
Hamiltonian matrix $H_{\nu \nu ^{\prime }}=\langle \nu |H|\nu ^{\prime
}\rangle $. Accordingly, the singularities are the branch points $\tau
_{c}(k),$ $k=1,2,\ldots $ of $E_{+}(\tau )-E_{-}(\tau )$. Introducing a new
variable \cite{10,22}
\begin{equation}
z(\tau )=\int\limits_{0}^{\tau }d\tau ^{\prime }[E_{+}(\tau ^{\prime
})-E_{-}(\tau ^{\prime })]  \label{eq51}
\end{equation}
it is shown in the Appendix A that after taking the limit $n\rightarrow
\infty $ the nonadiabatic transition probability is given by
\begin{equation}
P(\delta )\cong \exp \left[ -(F_{g}^{ns}+\frac{1}{\delta }F_{d}^{ns})\right]
\left| \sum\limits_{k}\exp F_{g}^{s}(k)e^{\frac{i}{\delta }z_{c}(k)}\right|
^{2}  \label{eq52}
\end{equation}
with the \textit{singular} ``geometrical'' contribution
\begin{equation}
F_{g}^{s}(k)=\int\limits_{0}^{z_{c}(k)}dz\frac{\frac{d\alpha _{+}}{dz}(z)-%
\frac{d\alpha _{-}}{dz}(z)}{\alpha _{+}(z)-\alpha _{-}(z)}  \label{eq53}
\end{equation}
and the singular points $z_{c}(k)=z(\tau _{c}(k))$, which are \textit{above}
the contour $\mathcal{C}=\{z(\tau )|-\infty \leq \tau \leq \infty \}$. The
final result of this section, Eq.~(\ref{eq52}), is the generalization of the
DDP formula (as it has been rigorously proven for hermitean TLS-Hamiltonians
\cite{10}) to nonhermitean ones, describing dissipative TLS. The reader
should note that the use of the superadiabatic basis leads to a
pre-exponential factor in Eq.~(\ref{eq52}) which is equal to one, which is
identical to the case without dissipation. The result, Eq.~(\ref{eq52}),
exhibits that the ``dynamical'' contributions follow from the adiabatic
eigenvalues and their branch points, whereas the ``geometrical''
contributions involve $\alpha _{\pm }(\tau )$, only. If we parametrize for a
TLS with hermitean Hamiltonian the external field components $B_{j}$, Eq.~(%
\ref{eq6}), as it has been done in Ref.~\cite{11}, one recovers that $%
F_{g}^{ns}=0$ and that Eq.~(\ref{eq53}) becomes:
\begin{equation}
F_{g}^{s}=\int\limits_{0}^{\tau _{c}(k)}d\tau \;\dot{\phi}(\tau )\cos \Theta
(\tau )  \label{eq54}
\end{equation}
in agreement with the result in Ref.~\cite{11}.

\begin{figure}[tbp]
\centerline{\includegraphics[width=8cm,clip]{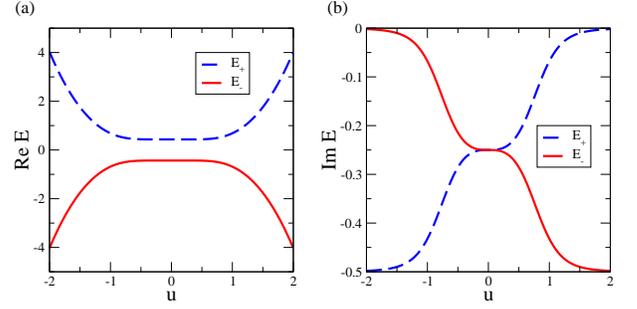}} \vspace*{8pt}
\caption{$u$-dependence of the adiabatic eigenvalues for a power law sweep $%
\tilde{w}(u)=u^{3}$ and $\tilde{\protect\gamma}=0.5<\tilde{\protect\gamma}%
_{c}=1$ (a) real part of $E_{\pm }(u)$, and (b) imaginary part of $E_{\pm
}(u)$ }
\label{fig1}
\end{figure}

\begin{figure}[tbp]
\centerline{\includegraphics[width=8cm,clip]{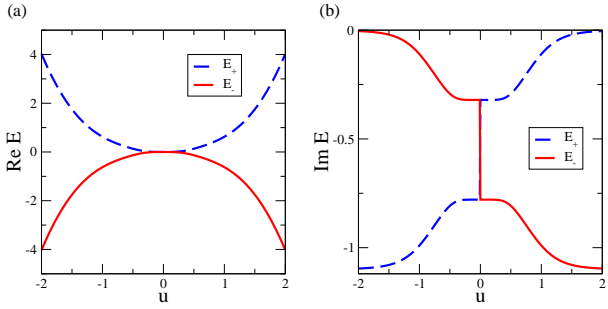}} \vspace*{8pt}
\caption{Same as Fig. 1, but for $\tilde{\protect\gamma}=1.1>\tilde{\protect%
\gamma}_{c}$ }
\label{fig2}
\end{figure}

\section{Application to the Akulin-Schleich Model}

The AS-model is given by \cite{19}
\begin{equation}
H(t)=-\frac{1}{2}\left[ W(t)\sigma _{z}+\Delta \sigma _{x}+i\gamma (\sigma
_{z}-\sigma _{0})\right]  \label{eq55}
\end{equation}
with the external field $W(t)$, the tunnelling matrix element $\Delta $, and
the damping constant $\gamma \geq 0$ of level $|2\rangle \hat{=}|\uparrow
\rangle $. $\sigma _{0}$ is the $2\times 2$ unit matrix. Let us introduce
dimensionless quantities:
\begin{equation}
\tilde{w}(u)=\frac{W(t)}{\Delta },\quad u=\frac{vt}{\Delta },\quad \tilde{%
\gamma}=\frac{\gamma }{\Delta },\quad \tilde{\epsilon}=\frac{\Delta ^{2}}{%
\hbar v}.  \label{eq56}
\end{equation}
Note that the time variable $\tau $ of the previous section is not
dimensionless. After the replacement of $\tau $ by $u$, Eq.~(\ref{eq1})
takes the form of Eq.~(\ref{eq8}) with:
\begin{equation}
\delta =\tilde{\epsilon}^{-1}.  \label{eq57}
\end{equation}
From Eqs.~(\ref{eq49}) and (\ref{eq50}) it follows immediately
\begin{equation}
E_{\pm }(u)=\frac{1}{2}\left[ -i\tilde{\gamma}\pm \sqrt{(\tilde{w}(u)+i%
\tilde{\gamma})^{2}+1}\right]  \label{eq58}
\end{equation}
\begin{equation}
\alpha _{\pm }(u)=-(\tilde{w}(u)+i\tilde{\gamma})\pm \sqrt{(\tilde{w}(u)+i%
\tilde{\gamma})^{2}+1},  \label{eq59}
\end{equation}
where the branch of the square root has been chosen such that $\sqrt{x}\geq
0 $, for $x\geq 0$. Fig.~\ref{fig1}a and Fig.~\ref{fig1}b exhibit $\mathrm{Re%
}E_{\pm }(u)$, and\textrm{\ }$\mathrm{Im}E_{\pm }(u)$, respectively, for an
analytical power law sweep $\tilde{w}(u)=u^{3}$ and for $\tilde{\gamma}<1$.
The corresponding result for $\tilde{\gamma}>1$ is shown in Figures \ref
{fig2}a and \ref{fig2}b. In the following we will consider crossing sweeps,
only. For those it is
\begin{equation}
\lim_{u\rightarrow \pm \infty }\tilde{w}(u)=\pm \infty .  \label{eq60}
\end{equation}
Returning sweeps for which $\lim_{u\rightarrow \pm \infty }\tilde{w}%
(u)=-\infty $ (or $+\infty )$ can be treated analogously. It is easy to
prove that
\begin{equation}
E_{\pm }(u)=\frac{1}{2}\left[ \tilde{w}(u)+\frac{1}{2\tilde{w}(u)}-\frac{i%
\tilde{\gamma}}{2\tilde{w}^{2}(u)}+\mathcal{O}(\tilde{w}^{-3}(u))\right]
\label{eq61}
\end{equation}
for $u\rightarrow \pm \infty $ and
\begin{equation}
E_{\pm }(u)=\frac{1}{2}\left[ -\tilde{w}(u)-2i\tilde{\gamma}-\frac{1}{2%
\tilde{w}(u)}+\mathcal{O}(\tilde{w}^{-2}(u))\right]  \label{eq62}
\end{equation}
for $u\rightarrow \mp \infty .$

From Eqs.~(\ref{eq61}) and (\ref{eq48}) it follows that $F_{d}^{ns}$ is
finite provided $\int\limits^{u}du^{\prime }\tilde{w}^{-3}(u^{\prime })$
exists for $u\rightarrow \pm \infty $. This is fulfilled if $\tilde{w}(u)$
decays faster than $u^{-1/3}$. Otherwise $F_{d}^{ns}=\infty $ which makes $%
P(\delta )$ to vanish. Figs. \ref{fig1} and \ref{fig2} demonstrate that
there exists a critical value for $\tilde{\gamma}_{c}$. For $0\leq \tilde{%
\gamma}<\tilde{\gamma}_{c}=1$ we have $\mathrm{Re}[E_{+}(u)-E_{-}(u)]>0$ for
all $u$ and $\mathrm{Im}E_{\sigma }(u)$ is continuous whereas $\mathrm{Re}%
[E_{+}(u)-E_{-}(u)]$ vanishes if $\tilde{w}(u)=0$ and $\mathrm{Im}E_{\sigma
}(u)$ becomes discontinuous, provided $E_{\pm }(u)$ are defined by Eq.~(\ref
{eq58}).

The nonsingular geometrical part, Eq.~(\ref{eq47}), can be calculated
without specifying the $u$-dependence of $\tilde{w}$. Substituting $\alpha
_\pm (u)$ and $\dot{\alpha}_\pm (u)$ from Eq.~(\ref{eq59}) into Eq.~(\ref
{eq47}), both integrals in Eq.~(\ref{eq47}), become a sum of two integrals.
One of them can be calculated by the introduction of a new integration
variable $\zeta = \tilde{w}+ i \tilde{\gamma}$ and the other by noticing
that its integrand can be rewritten as a derivative of a logarithm with
respect to $u$. Without restricting generality we assume that $\tilde{w}%
(0)=0 $. Then we obtain with Eq.~(\ref{eq60})

\begin{equation}
F_{g}^{ns}(\tilde{\gamma})=2\mathrm{Re}\left[ \ln (i\tilde{\gamma}+\sqrt{1-%
\tilde{\gamma}^{2}})\right] .  \label{eq63}
\end{equation}

\begin{figure}[tbp]
\centerline{\includegraphics[width=7cm,clip]{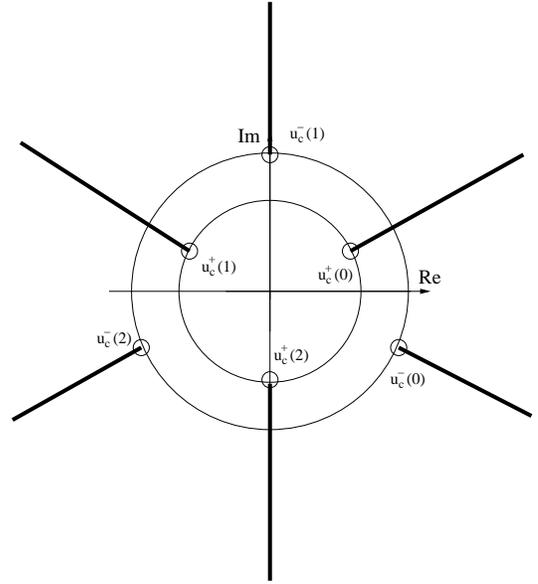}} \vspace*{8pt}
\caption{Branch points (open circles) $u_{c}^{\pm }(k)$ of $%
E_{+}(u)-E_{-}(u) $ for a power law sweep $\tilde{w}(u)=u^{3}$ and $\tilde{%
\protect\gamma}<\tilde{\protect\gamma}_{c}$. The thick solid lines are the
branch cuts. The radius of the inner and outer circle is $(1-\tilde{\protect%
\gamma})^{1/3}$ and $(1+\tilde{\protect\gamma})^{1/3}$, respectively }
\label{fig3}
\end{figure}

The nonsingular ``dynamical'' and both singular contributions require the
explicit $u$-dependence of $\tilde{w}$. As said above we will consider
crossing sweeps only. Therefore we restrict ourselves to power law sweeps $%
\tilde{w}(u)=u^{n}$ with $n>0$ and $n$ \textit{odd}. $n$ should not be
confused with the truncation number $n$ in the previous section. Since $%
\tilde{w}(-u)=-\tilde{w}(u)$ we can rewrite $F_{d}^{ns}$ as follows:
\begin{equation}
F_{d}^{ns}(\tilde{\gamma})=2\int\limits_{0}^{\infty }du\left[ \tilde{\gamma}-%
\mathrm{Im}\sqrt{(\tilde{w}(u)+i\tilde{\gamma})^{2}+1}\right] .  \label{eq64}
\end{equation}
It is easy to see that
\begin{equation}
F_{g}^{ns}(0)=0,\quad F_{d}^{ns}(0)=0,  \label{eq65}
\end{equation}
for $\tilde{\gamma}=0$. Hence, the nonsingular contributions to the
nonadiabatic transition probability vanish if there is no dissipation. In
this case the result (\ref{eq52}) reduces to that found by Berry \cite{11}
for hermitean Hamiltonians and for a single complex crossing point
contributing to Eq.~(\ref{eq52}). What remains is the determination of the
singular points $u_{c}(k)$, $k=1,2,\ldots $ and the calculation of $z_{c}(k)$
and $F_{g}^{s}(k)$. These singular points are the branch points of $%
E_{+}(u)-E_{-}(u)$. Their location depends on whether $0\leq \tilde{\gamma}<%
\tilde{\gamma}_{c}$ or $\tilde{\gamma}>\tilde{\gamma}_{c}=1$. Let us start
with the \textit{first} case $0\leq \tilde{\gamma}<\tilde{\gamma}_{c}$. From
$(u^{n}+i\tilde{\gamma})^{2}+1=0$, $n$ odd, we find
\begin{equation}
u_{c}^{\pm }(k)=\pm (1\mp \tilde{\gamma})^{1/\pi }\exp \left[ i\left( \frac{%
\pi }{2n}+k\frac{2\pi }{n}\right) \right]  \label{eq66}
\end{equation}
for $k=0,1,\ldots ,n-1,$ which are shown together with the branch cuts in
Figure 3 for $n=3$. From Eqs.~(\ref{eq51}) and (\ref{eq66}) we obtain the
corresponding singular points in the complex $z$-plane:
\begin{equation}
z_{c}^{\pm }(k)=\pm h_{n}^{\pm }(\tilde{\gamma})\exp \left[ i\left( \frac{%
\pi }{2n}+k\frac{2\pi }{n}\right) \right]  \label{eq67}
\end{equation}
where
\begin{equation}
h_{n}^{\pm }(\tilde{\gamma})=\int\limits_{0}^{(1\mp \tilde{\gamma})^{1/\pi
}}dx\sqrt{1-(\tilde{\gamma}\pm x^{n})^{2}}.  \label{eq68}
\end{equation}
Since the mapping $z(u)$ is analytic in the complex $u$-plane, except at the
branch lines, it is conformal. Accordingly, for those $u_{c}^{\pm }(k)$
which are in the upper $u$-plane the corresponding $z_{c}^{\pm }(k)$ will be
above the integration contour $\mathcal{C}$ and therefore will contribute to
$P$ (see end of the second section). After the determination of the singular
points we can proceed to calculate their ``geometrical'' and ``dynamical''
contribution to $P$. From Eqs.~(\ref{eq53}) and (\ref{eq59}) it follows:
\begin{eqnarray}
F_{g}^{s}(k) &=&\int\limits_{0}^{u_{c}^{\pm }(k)}du\frac{\dot{\alpha}_{+}(u)-%
\dot{\alpha}_{-}(u)}{\alpha _{+}(u)-\alpha _{-}(u)}  \nonumber  \label{eq69}
\\
&=&-\int\limits_{\tilde{w}(0)+i\tilde{\gamma}}^{\tilde{w}(u_{c}^{\pm }(k))+i%
\tilde{\gamma}}d\zeta \frac{1}{\sqrt{\zeta ^{2}+1}}  \nonumber \\
&=&-\ln (-\alpha _{-}(u_{c}^{\pm }(k)))+\ln (i\tilde{\gamma}+\sqrt{1-\tilde{%
\gamma}^{2}}).
\end{eqnarray}
Because $(\tilde{w}(u_{c}^{\pm }(k))+i\tilde{\gamma})+1=0$ we get from Eq.~(%
\ref{eq59}) that $\alpha _{-}(u_{c}^{\pm }(k))=1$ such that
\begin{equation}
F_{g}^{s}(k)=\ln (i\tilde{\gamma}+\sqrt{1-\tilde{\gamma}^{2}})+i\pi \equiv
F_{g}^{s}(\tilde{\gamma}).  \label{eq70}
\end{equation}
The reader should note that $F_{g}^{s}$ is independent on $k$. Consequently
it can be taken in front of the sum in Eq.~(\ref{eq52}) which yields $\exp (2%
\mathrm{Re}F_{g}^{s})$ and just cancels the non-singular ``geometrical''
factor $\exp (-2\mathrm{Re}F_{g}^{ns})$, due to Eq.~(\ref{eq63}). Therefore
we find that no ``geometrical'' factor occurs for the AS-model. This will
change if we apply an additional \textit{time-dependent} field in the $x$-
and $y$-direction. What remains is the calculation of the singular
``dynamical'' factor. Because we are interested in the adiabatic limit $%
\delta \rightarrow 0$, we have to take into account in Eq.~(\ref{eq52})
those singularities in the upper $z$-plane with smallest imaginary part.
These are $z_{c}^{+}(k=0)$ and $z_{c}^{+}(k=(n-1)/2)$, for which $\mathrm{Re}%
z_{c}^{+}(0)=-\mathrm{Re}z_{c}^{+}((n-1)/2)$ and (of course) $\mathrm{Im}%
z_{c}^{+}(0)=\mathrm{Im}z_{c}^{+}((n-1)/2)$. Using Eq.~(\ref{eq67}) with $%
k=0 $ and $k=(n-1)/2$ we obtain finally:
\begin{eqnarray}
P &\cong &4\cos ^{2}\left( \tilde{\epsilon}h_{n}^{+}(\tilde{\gamma})\cos
\frac{\pi }{2n}\right) \exp \left[ -\tilde{\epsilon}F_{d}^{ns}(\tilde{\gamma}%
)\right]  \nonumber \\
&&\times \exp \left[ -2\tilde{\epsilon}h_{n}^{+}(\tilde{\gamma})\right] \sin
\frac{\pi }{2n}.  \label{eq71}
\end{eqnarray}

\begin{figure}[tbp]
\centerline{\includegraphics[width=8cm,clip]{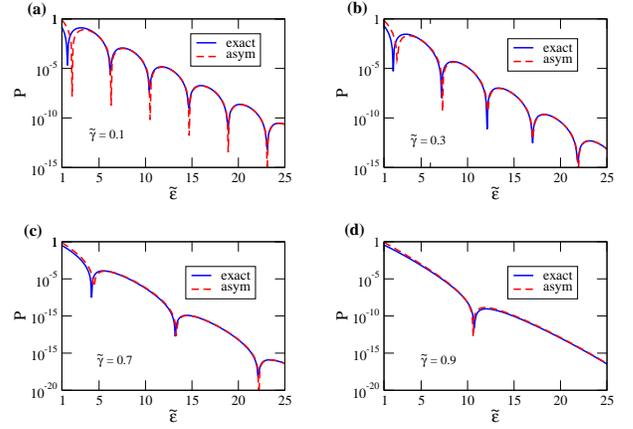}} \vspace*{8pt}
\caption{Comparison of the numerical exact (solid line) and the asymptotic
result, Eq.~(\ref{eq71}), (dashed line) for $P(\tilde{\protect\epsilon},%
\tilde{\protect\gamma})$ and a power law sweep $\tilde{w}(u)=u^{3}$. (a) $%
\tilde{\protect\gamma}=0.1$, (b) $\tilde{\protect\gamma}=0.3$, (c) $\tilde{%
\protect\gamma}=0.7$ and (d) $\tilde{\protect\gamma}=0.9$ }
\label{fig4}
\end{figure}

\begin{figure}[tbp]
\centerline{\includegraphics[width=8cm,clip]{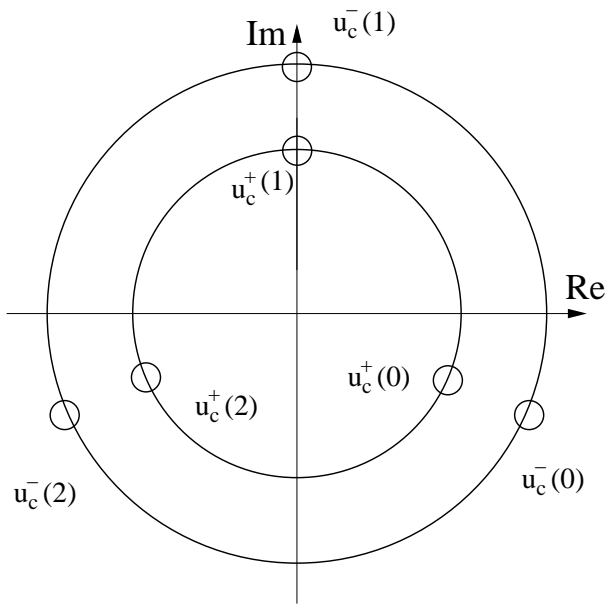}} \vspace*{8pt}
\caption{Same as Fig. 3 but for $\tilde{\protect\gamma}>\tilde{\protect\gamma%
}_{c}$ and without branch cuts. The radius of the inner and outer circle is $%
(\tilde{\protect\gamma}-1)^{1/3}$ and $(\tilde{\protect\gamma}+1)^{1/3}$. }
\label{fig5}
\end{figure}

\begin{figure}[tbp]
\centerline{\includegraphics[width=8cm,clip]{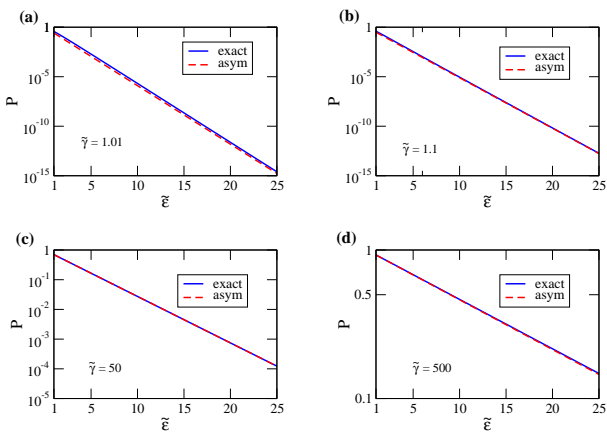}} \vspace*{8pt}
\caption{Same as Fig. 4, however for $\tilde{\protect\gamma}>\tilde{\protect%
\gamma}_{c}$ and with the asymptotic result, Eq.~(\ref{eq75}). (a) $\tilde{%
\protect\gamma}=1.01$, (b)$\tilde{\protect\gamma}=1.1$, (c) $\tilde{\protect%
\gamma}=50$ and (d) $\tilde{\protect\gamma}=500$ }
\label{fig6}
\end{figure}

\begin{figure}[tbp]
\centerline{\includegraphics[width=8cm,clip]{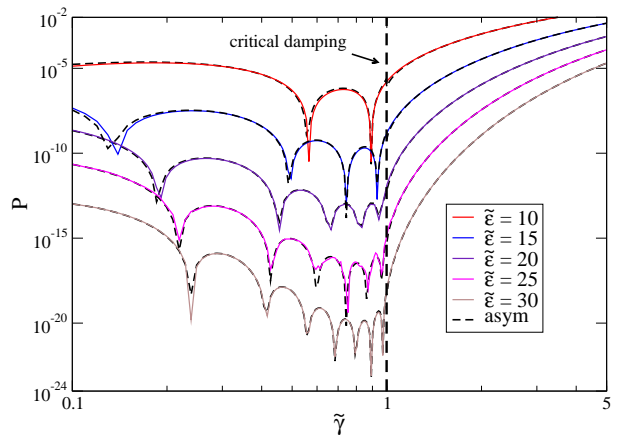}} \vspace*{8pt}
\caption{$\tilde{\protect\gamma}$-dependence of $P(\tilde{\protect\epsilon},%
\tilde{\protect\gamma})$ for $\tilde{w}(u)=u^{3}$ and $\tilde{\protect%
\epsilon}=10,15,20,25,30$ (from top to bottom). Numerical exact result
(solid line) and the asymptotic one (dashed line) }
\label{fig7}
\end{figure}

Let us consider linear sweeps, i.e. $n=1$. Then there exists only one
singularity $u_{c}^{+}(0)=i(1-\tilde{\gamma})$ in the upper $u$-plane and
Eq.~(\ref{eq52}) reduces to
\begin{equation}
P\cong \exp \left[ -\frac{1}{\delta }\left[ F_{d}^{ns}+2\mathrm{Im}%
z_{c}^{+}(0)\right] \right] .  \label{eq72}
\end{equation}
The exponent can be calculated analytically by using $u+i\tilde{\gamma}$ as
an integration variable. As a consequence one finds that the $\tilde{\gamma}$%
-dependence drops out from the exponent. With $\delta =\tilde{\epsilon}^{-1}$
one obtains
\begin{equation}
P\cong e^{-\epsilon }\;,\qquad \epsilon \equiv \frac{\pi }{2}\tilde{\epsilon}%
=\frac{\pi \Delta ^{2}}{2\hbar v},  \label{eq73}
\end{equation}
consistent with the finding in Ref.~\cite{19}. In order to check the
validity of Eq. (\ref{eq71}), we have solved numerically the time dependent
Schr\"{o}dinger equation in order to determine $P$. A comparison between the
numerically exact and the asymptotic result, Eq.~(\ref{eq71}), is shown in
Fig. \ref{fig4} for $n=3$ and four different $\tilde{\gamma}$-values. We
observe that the deviation between both results, e.g., for $\tilde{\epsilon}%
=5$ and $\tilde{\gamma}=0.3$, is about 1.6 per cent, only. Similarly good
agreement has been found for $n>3$. From Eq.~(\ref{eq71}) it follows that
there exist an infinite number of critical values $\tilde{\epsilon}%
_{c}^{(\nu )}(\tilde{\gamma}),\;\nu =1,2,3,\ldots $ at which the oscillatory
prefactor in Eq.~(\ref{eq71}) vanishes. From this we can conclude that these
St\"{u}ckelberg oscillations proven to exist for TLS without dissipation
\cite{10} and discussed later in Refs.~\cite{9,23} for $\tilde{\gamma}=0$
survive even in presence of dissipation, provided $\tilde{\gamma}<\tilde{%
\gamma}_{c}=1$. Indeed, we will see below that they disappear for $\tilde{%
\gamma}>\tilde{\gamma}_{c}$. It is not only the survival of the
oscillations, but also the survival of the \textit{complete} transitions
from state $|1\rangle \hat{=}|\downarrow \rangle $ to state $|2\rangle \hat{=%
}|\uparrow \rangle $ found in Refs.~\cite{9,23} for $\tilde{\gamma}=0$, as
long as $\tilde{\gamma}<\tilde{\gamma}_{c}$.

Now, we turn to the \textit{second} case $\tilde{\gamma}>\tilde{\gamma}_{c}$%
. For this case we find:
\begin{equation}
u_{c}^{\pm }(k)=-(\tilde{\gamma}\mp 1)^{^{1/n}}\exp \left[ i\left( \frac{\pi
}{2n}+k\frac{2\pi }{n}\right) \right] \;  \label{eq74}
\end{equation}
for $k=0,1,\ldots ,n-1,$ which are shown in Fig. \ref{fig5} for $n=3$. The
main difference to the case $0\leq \tilde{\gamma}<\tilde{\gamma}_{c}$ is
that there is exactly one singular point among $u_{c}^{\pm }(k)$ denoted by $%
u_{c}^{0}$ for which $z_{c}^{0}=z(u_{c}^{0})$ is on the real axis in the
complex $z$-plane. Using the definition, Eq.~(\ref{eq58}), of $E_{\pm }(u)$,
Fig. \ref{fig2} demonstrates that $E_{\pm }(u)$ is discontinuous on the real
u axis. There seem to exist two possibilities to deal with this problem.
First, after having chosen the branch cuts in the complex $u$-plane one has
to deform the integration contour along the real $u$-axis sucht that $u=0$
is above that contour and that no branch cut is crossed. This kind of
reasoning was used by Moyer in Ref.~\cite{20}. Second, one could define $%
E_{\pm }(u)$ such that they are analytic in a strip around the real $u$%
-axis. This can be done by interchanging $E_{+}(u)$ and $E_{-}(u)$ for $%
u\leq 0$. This has the consequence that the contour $z(u)$ for $u$ real is
in the right complex $z$-plane, starting e.g. above the positive real axis
for $u=-\infty $ , going through $z=0$ for $u=0$ and then continuing below
the positive real axis for $u\rightarrow +\infty $. This contour would
enclose $z_{0}$ if $\mathrm{Re}z_{0}>0$. Whether it can be closed such that
the closure does not make a contribution is not obvious. Since we are not
sure how to solve this problem in a rigorous manner, we have assumed that $%
z_{0}$ is the leading contribution to $P$, Eq.~(\ref{eq52}), for $\delta
\rightarrow 0$. Since $|e^{iz_{c}^{0}/\delta }|=1$ and due to the absence of
a ``geometrical'' contribution we obtain:
\begin{equation}
P\cong \exp \left[ -\tilde{\epsilon}F_{d}^{ns}(\tilde{\gamma})\right] ,
\label{eq75}
\end{equation}
with $F_{d}^{ns}\left( \tilde{\gamma}\right) $ given by Eq.~(\ref{eq64}). A
comparison between the $\tilde{\epsilon}$-dependence of the numerically
exact and the asymptotic result, Eq.~(\ref{eq75}), is presented in Fig. \ref
{fig6}. Again we find a very good agreement already for $\tilde{\epsilon}%
\geq 1$. This strongly supports the correctness of our assumption that $%
z_{0} $ is the most important singularity. Eq.~(\ref{eq75}) reveals that the
St\"{u}ckelberg oscillations as function of $\tilde{\epsilon}$ have
disappeared. We stress that both asymptotic results, Eq.~(\ref{eq71}) and (%
\ref{eq75}), are valid for all $\tilde{\gamma}$ with $0\leq \tilde{\gamma}<%
\tilde{\gamma}_{c}$ and for all $\tilde{\gamma}$ larger than $\tilde{\gamma}%
_{c}$, respectively, provided $\tilde{\epsilon}$ is large enough. This is
demonstrated in Fig. \ref{fig7} for different $\tilde{\epsilon}$.

\begin{figure}[tbp]
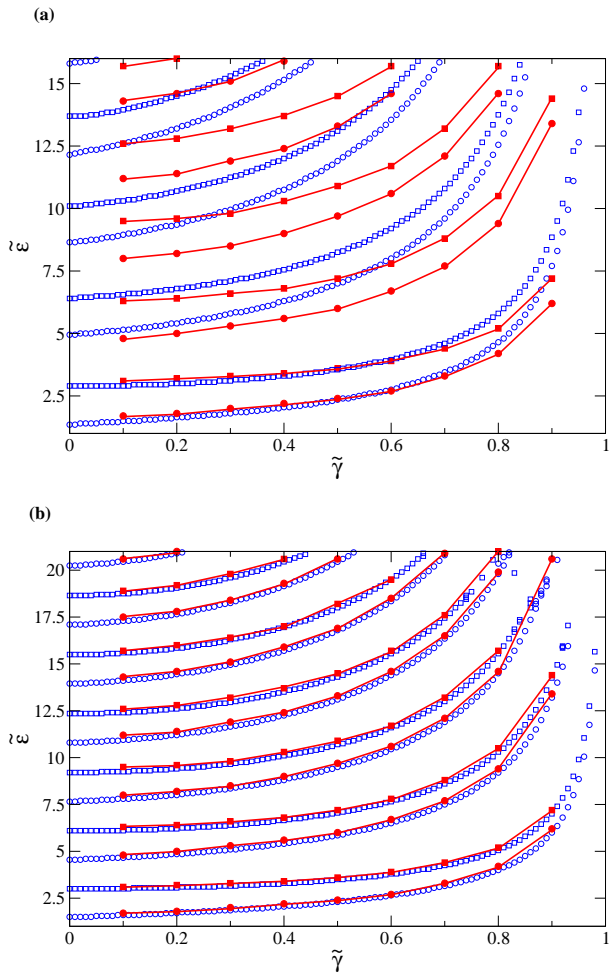

\centerline{\includegraphics[width=8cm,clip]{fig_8a.eps}} \vspace*{8pt}%
\centerline{\includegraphics[width=8cm,clip]{fig_8b.eps}} \vspace*{8pt}
\caption{Comparison of the critical values $\tilde{\protect\epsilon}_{c}^{(%
\protect\nu )}(\tilde{\protect\gamma})$ for (a) $\tilde{w}(u)=u^{5}$ and (b)
$\tilde{w}(u)=u^{51}$. The numerically exact result is shown by the open
circles and the result obtained for the oscillator model is depicted by the
full circles. The solid lines are a guide for the eye. }
\label{fig8}
\end{figure}

\section{Interpretation by a Damped Harmonic Oscillator}

In this section we will give an intuitive explanation of the St\"{u}ckelberg
oscillations and will present an approximate calculation for the critical
values $\tilde{\epsilon}_{c}^{(\nu )}(\tilde{\gamma})$ for power law
crossing sweeps $\tilde{w}(u)=u^{n}$, $n$ odd. Close to the resonance at $%
u=0 $ we may neglect $\tilde{w}(u)$. Then the time dependent Schr\"{o}dinger
equation for the amplitude of state $|1\rangle $
\begin{equation}
\tilde{c}_{1}(u)=c_{1}(u)\exp \left[ i\int\limits_{-\infty }^{u}du^{\prime }%
\tilde{w}(u^{\prime })\right]  \label{eq76}
\end{equation}
becomes
\begin{equation}
\ddot{\tilde{c}}_{1}+2\mu \dot{\tilde{c}}_{1}(u)+\omega _{0}^{2}\tilde{c}%
_{1}(u)\cong 0  \label{eq77}
\end{equation}
with:
\begin{equation}
\mu =\frac{\tilde{\epsilon}\tilde{\gamma}}{2},\qquad \omega _{0}=\frac{%
\tilde{\epsilon}}{2}.  \label{eq78}
\end{equation}
Let $t_{\text{trans}}$ be the Landau-Zener transition time. In the adiabatic
limit it is well-known that $t_{\text{trans}}=\Delta /v$. Eq.~(\ref{eq56})
yields $u_{\text{trans}}=1$. Therefore we will require as initial
conditions:
\begin{eqnarray}
\tilde{c}_{1}(-u_{\text{trans}}=-1) &=&1  \nonumber  \label{eq79} \\
\dot{\tilde{c}}_{1}(-u_{\text{trans}}=-1) &=&0.
\end{eqnarray}
Eq.~(\ref{eq77}) is the equation of motion for a damped harmonic oscillator
which can easily be solved. The special solutions are $\exp \left[ i\omega
_{\pm }(\tilde{\epsilon},\tilde{\gamma})u\right] $ with
\begin{equation}
\omega (\tilde{\epsilon},\tilde{\gamma})=\frac{\tilde{\epsilon}}{2}\left[ i%
\tilde{\gamma}\pm \sqrt{1-\tilde{\gamma}^{2}}\right] .  \label{eq80}
\end{equation}
This result makes obvious the existence of a critical damping $\tilde{\gamma}%
_{c}=1$. For $0\leq \tilde{\gamma}<\tilde{\gamma}_{c}$ and $\tilde{\gamma}<%
\tilde{\gamma}_{c}$ the oscillator is underdamped and overdamped,
respectively. This qualitative different behavior is the origin of the
different $\tilde{\epsilon}$-dependence of $P$ for $0\leq \tilde{\gamma}<%
\tilde{\gamma}_{c}$ and $\tilde{\gamma}<\tilde{\gamma}_{c}$, found in the
third section. This relationship can be deepened more by calculating $\tilde{%
\epsilon}_{c}^{(\nu )}(\tilde{\gamma})$. Having solved Eq.~(\ref{eq77}) with
initial conditions, Eq.~(\ref{eq79}) we approximate $P$ by:
\begin{equation}
P\cong |c_{1}(+u_{\text{trans}})=+1|^{2}=|\tilde{c}_{1}(+u_{\text{trans}%
}=+1)|^{2}.  \label{eq81}
\end{equation}
The zeros (with respect to $\tilde{\epsilon}$) of $P$ yield $\tilde{\epsilon}%
_{c}^{(\nu )}(\tilde{\gamma})$. A numerical solution of the corresponding
transcendental equation leads to the results shown in Figure 8 for $\tilde{w}%
(u)=u^{n}$ with $n=5$ and $n=51$ and $\tilde{\gamma}<\tilde{\gamma}_{c}$.
Figure 8 also contains the result from a numerically exact solution of the
time dependent Schr\"{o}dinger equation. Comparing both results we observe
that the agreement for $n=5$ is qualitatively good, but quantitatively less
satisfactory. However, increasing $n$ more and more leads even to a rather
good quantitative agreement, as can be seen for $n=51$. This behavior is
easily understood, since $\tilde{w}(u)$ within the transition range $(-1,1)$
becomes practically zero for $n$ large enough. Figure 8 also demonstrates
that $\tilde{\epsilon}_{c}^{(\nu )}$ increases monotonically with $\tilde{%
\gamma}$ which is related to the decrease of $\mathrm{Re}\omega (\tilde{%
\epsilon},\tilde{\gamma})$ for increasing $\tilde{\gamma}$. The oscillator
model can also be used to determine a lower bound for $\tilde{\epsilon}%
_{c}^{(1)}(\tilde{\gamma}=0)$. For $u_{\text{trans}}=1$ one gets
\begin{equation}
\tilde{\epsilon}_{c}^{(1)}(\tilde{\gamma}=0)\geq \frac{\pi }{2}  \label{eq82}
\end{equation}
such that $\tilde{\epsilon}_{c}^{(\nu )}(\tilde{\gamma})\geq \tilde{\epsilon}%
_{c}^{(1)}(\tilde{\gamma})>\tilde{\epsilon}_{c}^{(1)}(\tilde{\gamma}=0)\geq
\pi /2$, for all $\tilde{\gamma}$. It is interesting that the lower bound (%
\ref{eq82}) for $\tilde{\epsilon}$ is similar to that obtained from the
\textit{inverse} Landau-Zener problem \cite{25}. There, the $t$-dependent
survival probability $P(t;\tilde{\epsilon})$ is given and $W(t;\tilde{%
\epsilon})$ is determined analytically from $P(t;\tilde{\epsilon})$. If $P(t;%
\tilde{\epsilon})=P(t;\tilde{\epsilon}u)$, with $u$ and $\tilde{\epsilon}$
from Eq.~(\ref{eq56}), varies from one (for $t=-\infty )$ to zero (for $%
t=+\infty )$, it is found that a solution $W(t;\tilde{\epsilon})$ of the
inverse problem only exists, if
\begin{equation}
\tilde{\epsilon}>1.  \label{eq83}
\end{equation}
The latter inequality, as well as inequality (\ref{eq82}) implies that the
ratio $t_{\text{trans}}/t_{\text{tunnel}}$ of the transit time $t_{\text{%
trans}}=\Delta /v$ and the time period of coherent tunneling $t_{\text{tunnel%
}}=\hbar /v$, which equals $\tilde{\epsilon}$, is of order one. It is
obvious that \textit{complete} transitions can not occur if $t_{\text{trans}%
} $ is too small compared to $t_{\text{tunnel}}$, i.e. for $\tilde{\epsilon}%
\ll 1$. In that case the quantum system does not have time enough to tunnel
from the initial state $|1\rangle $ to state $|2\rangle $.

\section{Summary and Conclusions}

Our main focus has been on the derivation of the non-adiabatic transition
probability $P(\tilde{\epsilon})$ for a dissipative two-level system
modelled by a \textit{general} non-hermitean Hamiltonian, depending
analytically on time. Following for the hermitean case Berry's approach by
use of a superadiabatic basis we have found a generalization of the
DDP-formula. Besides a ``geometrical`` and a ``dynamical`` factor,
completely determined by the crossing points in the complex time plane, we
also have found a non-universal ``geometrical'' and ``dynamical``
contribution to $P$. The latter require the knowledge of the Hamiltonian's
full time dependence and are identical to zero in the absence of
dissipation. Without specification of the TLS-Hamiltonian, we have shown
that both ``geometrical'' contributions can be expressed by $\alpha _\pm (u)$%
, the ratio of the components of each adiabatic states $|u_{0,\pm}(u)\rangle$
in the basis $|\nu\rangle,\,\,\nu=1,2$, and both ``dynamical'' ones by the
adiabatic eigenvalues $E_\pm (u)$, only. In this respect our result for $P(%
\tilde{\epsilon})$ is independent of a special parametrization of the
Hamiltonian matrix. Although the result in Ref.~\cite{20} is not in such an
explicit form like Eq.~(\ref{eq46}) the existence of this nonsingular
``dynamical'' contribution has already been stated there. However, the
nonsingular ``geometrical'' part, Eq.~(\ref{eq47}), has not been found in
that paper. \newline
As a physical application we have studied the AS-model \cite{19}. This model
describes a dissipative TLS where the initial upper level is damped. In \cite
{19} it has been shown that the probability $P$ for a linear time dependence
of the bias does not depend on the damping constant $\tilde{\gamma}$ for all
$\tilde{\epsilon}$. Our results demonstrate that this is not generic. For
instance, nonlinear power law crossing sweeps generate a $\tilde{\gamma}$%
-dependence of $P$. For such sweeps a critical value $\tilde{\gamma}_c=1$
exists. Below $\tilde{\gamma}_c$ the non-adiabatic transition probability
oscillates and vanishes at critical values $\tilde{\epsilon}_c^{(\nu)}(%
\tilde{\gamma})$, and for $\tilde{\gamma} > \tilde{\gamma}_c$ the
oscillations are absent. Hence, the existence of complete transitions at an
infinite set of critical sweep rates still holds for all $\tilde{\gamma}$
below $\tilde{\gamma}_c$. In the section IV we have shown how the
oscillations and their disappearance for $\tilde{\gamma} > \tilde{\gamma}_c$
can be qualitatively explained by a damped harmonic oscillator. For power
law sweeps with rather large exponent, e.g. $n=50$, this description becomes
even quantitatively correct. No doubt, it would be interesting to study a
microscopic model of a TLS coupled to phonons, e.g. a spin-boson-Hamiltonian
as in Ref.~\cite{13}, in order to check whether the $\tilde{\epsilon}$%
-dependence of $P$ exhibits oscillations for power law sweep with $n>1$ and
small enough spin-phonon coupling. Another question concerns the interaction
between the TLS which have been completely neglected in our present work.
That they can play a crucial role was shown recently \cite{26}. Whether the
oscillations still exist in the presence of interactions between the TLS is
not obvious.

\begin{acknowledgements}We gratefully acknowledge discussions with
A. Joye and V. Bach. \end{acknowledgements}

\section*{Appendix}

\appendix

In this appendix we will describe how the asymptotic result, Eq.~(\ref{eq52}%
), has been derived from (\ref{eq46}). Although we follow Berry's approach
\cite{11,22} we repeat the most important steps since the non-hermitean
property of $H$ does not allow the simple parametrization used in Ref.~\cite
{11} and is not of the form of Eq.~(\ref{eqA1}) or Eq.~(\ref{eqA4}) of Ref.~
\cite{22}. Nevertheless we will recover the same universal recursion
relation for the coefficients $a_{m}^{-}(\tau )$ as found by Berry \cite{22}%
. In order to show how Eq.~(\ref{eq52}) can be obtained from Eq.~(\ref{eq46}%
) we have to calculate the three pre-exponential factors $E_{+}(\tau
)-E_{-}(\tau ),[\alpha _{+}(\tau )-\alpha _{-}(\tau )]/\dot{\alpha}_{+}(\tau
)$ and $\dot{a}_{n+1}^{-}(\tau )$ in Eq.~(\ref{eq46}). We assume that the
Hamiltonian $H(\tau )$ is analytic in $\tau $. Let $\tau _{c}$ be one of the
branch points of $E_{+}(\tau )-E_{-}(\tau )$. Close to $\tau _{c}$ we get:
\begin{equation}
E_{+}(\tau )-E_{-}(\tau )\cong c(\tau -\tau _{c})^{1/2}  \label{eqA1}
\end{equation}
with $c$ a constant, depending on $\tau _{c}$. Eq.~(\ref{eq51}) implies
\begin{equation}
z-z_{c}\cong \frac{2}{3}c(\tau -\tau _{c})^{3/2},  \label{eqA2}
\end{equation}
where $z_{c}=z(\tau _{c})$. In the adiabatic limit $\delta \rightarrow 0$
the main contribution to the integral (2.~line of Eq.~(\ref{eq46})) comes
from the singular points $\tau _{c}$ and $z_{c}$, respectively. Consequently
we have to calculate the pre-exponential factors (2. line of Eq.~(\ref{eq46}%
)) close to the singularities, only. Let us start with $[\alpha _{+}(\tau
)-\alpha _{-}(\tau )]/\dot{\alpha}_{+}(\tau )$. Using Eq.~(\ref{eq50}) it
follows with $\alpha _{\pm }(z)=\alpha _{\pm }(\tau (z))$ close to $z_{c}$
\begin{equation}
\lbrack \alpha _{+}(z)-\alpha _{-}(z)]/\alpha _{+}^{\prime }(z)\cong
6(z-z_{c}),  \label{eqA3}
\end{equation}
where $^{\prime }$ denotes derivative with respect to $z$. Note that $%
H_{11},H_{12}$ and $H_{22}$ do not enter in Eq.~(\ref{eqA3}). The
calculation of $\dot{a}_{n+1}^{-}(\tau )$ is more evolved. As a first step
we eliminate $\dot{b}_{m-1}^{-}(\tau )$, $b_{m-1}^{-}(\tau )$ and $%
b_{m}^{-}(\tau )$ from Eqs.~(\ref{eq23}), (\ref{eq24}) which yields a
recursion relation for $a_{m}^{-}(\tau )$:
\begin{eqnarray}
\dot{a}_{m}^{-}(\tau ) &=&\frac{i}{E_{+}(\tau )-E_{-}(\tau )}\left[ \ddot{a}%
_{m-1}^{-}(\tau )-\frac{\dot{\kappa}_{-}(\tau )}{\kappa _{-}(\tau )}\dot{a}%
_{m-1}^{-}(\tau )\right.  \nonumber \\
&&\qquad -\left. \kappa _{-}(\tau )\kappa _{+}(\tau )a_{m-1}^{-}(\tau )
\right] .  \label{eqA4}
\end{eqnarray}
Next we calculate the various terms close to $\tau _{c}$. From Eqs.~(\ref
{eq30}) and (\ref{eq50}) we get:
\begin{equation}
\frac{\dot{\kappa}_{-}(\tau )}{\kappa _{-}(\tau )}\cong (\tau -\tau
_{c})^{-1}  \label{eqA5}
\end{equation}
and
\begin{equation}
\kappa _{-}(\tau )\kappa _{+}(\tau )\cong \frac{1}{16}(\tau -\tau _{c})^{-2}.
\label{eqA6}
\end{equation}
Expressing the $\tau $-derivatives of $a_{m}^{-}$ and $a_{m-1}^{-}$ by
derivatives with respect to $z$:
\begin{equation}
\dot{a}_{m}^{-}(\tau )\cong c(\tau -\tau _{c})^{1/2}a_{m}^{-^{\prime }}(z)
\label{eqA7}
\end{equation}
and
\begin{equation}
\ddot{a}_{m}(\tau )\cong c^{2}(\tau -\tau _{c})a_{m}^{-^{\prime \prime }}(z)+%
\frac{c}{2}(\tau -\tau _{c})^{-1/2}a_{m}^{-^{\prime }}(z),  \label{eqA8}
\end{equation}
where $dz/d\tau =E_{+}(\tau )-E_{-}(\tau )$ and (\ref{eqA1}) was used, we
get from Eq.~(\ref{eqA4}) with Eq.~(\ref{eqA5}), (\ref{eqA6}):
\begin{equation}
a_{m}^{-^{\prime \prime }}(z)\cong (-i)\left[ \frac{a_{m-1}^{-}(z)}{%
36(z-z_{c})^{2}}-\frac{a_{m-1}^{-^{\prime }}(z)}{(z-z_{c})}%
-a_{m-1}^{-^{\prime \prime }}(z)\right]  \label{eqA9}
\end{equation}
with initial condition (cf. Eq.~(\ref{eq28})):
\begin{equation}
a_{0}^{-}(z)\equiv 1.  \label{eqA10}
\end{equation}
The recursion relation is identical to Eq.~(\ref{eq30}) in Ref.~\cite{22},
except the different sign in front of the square bracket. The sign change is
irrelevant. The exact solution of Eq.~(\ref{eqA9}), (\ref{eqA10}) can be
taken from Ref.~\cite{22}:
\begin{equation}
a_{m}^{-}(z)\cong B_{m}(z-z_{c})^{-m}  \label{eqA11}
\end{equation}
with
\begin{equation}
B_{m}=i^{m}\frac{(m-\frac{7}{6})!(m-\frac{5}{6})!}{m!(-\frac{7}{6})!(-\frac{5%
}{6})!}.  \label{eqA12}
\end{equation}
Then we get from Eq.~(\ref{eq46}) with (\ref{eqA1}) - (\ref{eqA3}) and (\ref
{eqA11}), (\ref{eqA12})
\begin{eqnarray}
&&\int\limits_{-\infty }^{\infty }d\tau \;\dot{a}_{n+1}^{-}(\tau )\frac{%
\alpha _{-}(\tau )-\alpha _{-}(\tau )}{\dot{\alpha}_{+}(\tau )}[E_{+}(\tau
)-E_{-}(\tau )]I(\tau )  \nonumber  \label{eqA13} \\
&=&\int\limits_{\mathcal{C}}dz\;a_{n+1}^{-^{\prime }}(z)\frac{\alpha
_{+}(z)-\alpha _{-}(z)}{\alpha _{+}^{\prime }(z)}I(z)  \nonumber \\
&=&-6(n+1)B_{n+1}\sum\limits_{k}\int\limits_{\mathcal{C}}dz\frac{1}{%
(z-z_{c}(k))^{n+1}}I(z)  \nonumber \\
&=&-6(n+1)2\pi i\left( \frac{i}{\delta }\right)
^{n}B_{n+1}\sum\limits_{k}I(z_{c}(k)),
\end{eqnarray}
where the sum over $k$ is restricted to all singular points $z_{c}(k)$ above
the contour $\mathcal{C}=\{z=z(\tau )|-\infty \leq \tau \leq \infty \}$.
Since
\begin{equation}
-6\cdot 2\pi (n+1)B_{n+1}/n!\longrightarrow i^{n+1},\quad n\rightarrow
\infty ,
\end{equation}
the prefactor in front of $\sum\limits_{k}$ in Eq.~(\ref{eqA13}) equals $%
(-1)^{n+1}$. Substituting (\ref{eqA13}) into Eq.~(\ref{eq46}) $\delta ^{n}$
cancels and one obtains the result Eq.~(\ref{eq52}).


\begin{thebibliography}{99}
\bibitem{1}  L. D. Landau, Phys. Z. \textit{Sowjetunion} \textbf{2}, 46
(1932)

\bibitem{2}  C. Zener, R. Soc. London, Ser. A \textbf{137}, 696 (1932)

\bibitem{3}  E. C. G. St\"uckelberg, Helv. Phys. Acta \textbf{5}, 369 (1932)

\bibitem{4}  E. Majorana, Nuovo Cimento \textbf{9}, 43 (1932)

\bibitem{5}  A. M. Dykhne, Sov. Phys. JETP \textbf{14}, 941 (1962)

\bibitem{6}  V. L. Pokrovskii, S. K. Savvinykh and F. R. Ulinich, Sov. Phys.
JETP \textbf{34}, 879 (1958)

\bibitem{7}  J. P. Davis and P. Pechukas, J. Chem. Phys. \textbf{64}, 3129
(1976)

\bibitem{8}  T. F. George and Y.-W. Lin, J. Chem. Phys. \textbf{60}, 2340
(1974)

\bibitem{9}  N. V. Vitanov and K.-A. Suominen, Phys. Rev. A\textbf{59}, 4580
(1999)

\bibitem{10}  A. Joye, G. Mileti and Ch.-Ed. Pfister, Phys. Rev. A\textbf{44}%
, 4280 (1991)

\bibitem{11}  M. V. Berry, Proc. R. Soc. Lond. \textbf{A430}, 405 (1990)

\bibitem{12}  A. Joye, H. Kunz and Ch.-Ed. Pfister, Ann. Phys. \textbf{208},
299 (1991)

\bibitem{13}  P. Ao and J. Rammer, Phys. Rev. Lett. \textbf{62}, 3004
(1989); Phys. Rev. \textbf{B43}, 5397 (1991)

\bibitem{14}  Y. Kayanuma, J. Phys. Soc. Jpn., \textbf{53}, 108 (1984)

\bibitem{15}  K. Saito and Y. Kayanuma, Phys. Rev. \textbf{A65}, 033407
(2002)

\bibitem{16}  M. Nishino, K. Saito and S. Miyashita, Phys. Rev \textbf{B65},
014403 (2001)

\bibitem{17}  V. L. Pokrovsky and N. A. Sinitsyn, Phys. Rev. B\textbf{67},
144303 (2003); \textbf{B69}, 104414 (2004)

\bibitem{18}  V. L. Prokovsky and S. Scheidl, Phys. Rev. \textbf{B70},
014416 (2004)

\bibitem{19}  V. M. Akulin and W. P. Schleich, Phys. Rev. \textbf{A46}, 4110
(1992)

\bibitem{20}  C. A. Moyer, Phys. Rev. \textbf{A64}, 033406 (2001)

\bibitem{21}  J. C. Garrison and E. M. Wright, Phys. Lett. \textbf{A128},
177 (1988)

\bibitem{22}  M. V. Berry, Proc. R. Soc. London \textbf{A429}, 61 (1990)

\bibitem{23}  D. A. Garanin and R. Schilling, Phys. Rev. \textbf{B66},
174438 (2002)

\bibitem{24}  M. V. Berry, Proc. R. Soc. London \textbf{A392}, 45 (1984)

\bibitem{25}  D. A. Garanin and R. Schilling, Europhys. Lett. \textbf{59}, 7
(2002)

\bibitem{26}  D. A. Garanin and R. Schilling, Phys. Rev. \textbf{B71},
184414 (2005)
\end{thebibliography}
\end{document}